\documentstyle[aps,epsf,prb
  ]{revtex}\input {amssym.def}
\begin{document}

\title{Magnetic Impurities in d-wave Superconductors}
\author{M.E. Simon, and C. M. Varma}
\address{Bell Laboratories, Lucent Technologies \\
Murray Hill, NJ 07974}

\maketitle

\begin{abstract}
We solve the problem of a magnetic impurity in a $d_{x^2-y^2}$-wave 
superconductor by a
variational method. A moment is found to exist in the superconducting state 
only if the Kondo-temperature in the normal state is larger than the 
maximum in the superconducting gap-function.
If a moment exists in the superconducting 
state, and provided spin-orbit coupling is non-zero, it induces a 
time-reversal breaking superconducting state locally around the inpurity 
which is
 a linear combination of the $d_{x^2-y^2}$ and $d_{xy}$ states. The 
 current pattern around the impurity in this state are evaluated.
\end{abstract}


\section*{Introduction}

Since magnetic impurities break time-reversal invariance, they tend to 
destroy superconductivity\cite{maki}. In superconducting ground states 
paired
 in a finite angular momentum (or its appropriate generalisation in a lattice), 
non-magnetic impurities are also pair-breaking since
 their potential, in general, does not transform in the same way as the 
order  parameter \cite{haas,har}. 

Recently a new phenomenon associated with magnetic
impurities in a $d_{x^{2}-y^{2}}$ condensate was proposed. It was argued
that the interaction between the magnetic impurity spin and the 
orbital moment of the condensate can help to stabilize a new        
 time-reversal violating phase:
$\alpha d_{x^{2}-y^{2}}$+$i\beta d_{xy}$\cite{bal1}. The physical point 
 is that such a state has a finite orbital moment around the impurity. 
Provided the spin-orbit coupling is finite, such a state interacts linearly     
with the magnetic moment. Therefore it is necessarily induced with the ratio 
$\beta /\alpha$ of order $E_{so}/\delta E_c$ where $E_{so}$ is the
 spin-orbit coupling energy and  $\delta E_c$ is the difference in 
condensation energy of the  $d_{x^{2}-y^{2}}$ and the $d_{xy}$ states. 

In this paper we examine this idea through a variational method introduced  
 for the Kondo (and mixed-valence) problem in normal metals \cite{var}. 
The variational method foreshadowed the development of the non-crossing
 approximation \cite{kei}, the 1/N method  \cite{ram} and the 
Slave-Boson approach \cite{col}. 
For the ground state properties these methods produce the same results 
as the variational approach.
First we ask the question: Under what condition can a moment exist in the
superconducting state. This is the question of the disappearance of the 
Kondo-effect in a d-wave superconductor. This problem also has been 
examined \cite{cas,ing}. 
We hope , our simple approach, with results equivalet to those derived earlier,
 makes the physical issues clearer. Because the density of states of
 quasi-particles goes to zero linearly in energy, the logarithmic Kondo 
singularity in the scattering matrix of the moment with conduction electrons   
is absent but, with a finite exchange coupling, the Kondo-effect and 
the disappearance of the magnetic moment is still possible. The condition 
for the moment to exist in the superconducting state is found to be 
$T_K/\Delta \lesssim 1$, where $T_K$ is the Kondo temperature in the normal
 state and $\Delta$ is the maximum in the superconducting gap-function.
(This is a more stringent condition than in s-wave superconductors.
\cite{sim,sato}). Second, we find the necessary condition
for inducing the locally time-reversal breaking state. In agreement with 
earlier conclusions, a finite spin-orbit coupling is found essential.
 With the variational wave-function, we also calculate the current 
distribution induced around the impurity. We also briefly discuss the 
problem of several impurities to conclude that a global 
time-reversal breaking phase 
is unlikely to result from this mechanism.

\section*{The Model}

We consider the usual BCS Hamiltonian ($H_{BCS}$) and the Anderson
Hamiltonian ($H_{I}$) to describe the superconductor and the impurity 
respectively:

\begin{equation}
H_{BCS}=\sum_{\bf{k}\sigma }\lambda _{\bf{k}}\gamma
_{\bf{k}%
\sigma }^{\dagger }\gamma _{\bf{k}\sigma }+E_{G}
\end{equation}

with $\lambda _{\bf{k}}=\sqrt{\epsilon _{\bf{k}}^{2}+\Delta _{%
\bf{k}}^{2}}$ , where $\epsilon _{\bf{k}}$ are the energy of
conduction electrons with respect to the chemical potential, $\Delta_{\bf{k}%
}=\sum_{\bf{k}^{\prime }}V_{kk^{\prime }}^{BCS}\langle c_{-\bf{k}%
\downarrow }c_{\bf{k}\uparrow }\rangle $ is the superconducting order
parameter, $V_{kk^{\prime }}^{BCS}$ is the BCS attractive interaction and
 $%
c_{\bf{k}\uparrow }^{\dagger }$ creates an electron with momentum $%
\bf{k}$ and spin $\uparrow $.
\begin{equation}
\gamma _{\bf{k}\uparrow }=u_{\bf{k}}c_{\bf{k}\uparrow}-v_{%
\bf{k}}c_{-\bf{k}\downarrow }^{\dagger }$, $\gamma_{-\bf{k}%
\downarrow }=v_{\bf{k}}c_{\bf{k}\uparrow }^{\dagger}+u_{\bf{k}%
}c_{-\bf{k}\downarrow} \label{bov}
\end{equation}
 annihilate and create the quasiparticles in the superconducting state and
\begin{equation}
u_{\bf{k}}=\sqrt{\frac{1}{2}(1+\frac{\epsilon _{\bf{k}}}{\lambda_{\bf{
k}}})},
 v_{\bf{k}}=\sqrt{\frac{1}{2}(1-\frac{\epsilon _{\bf{k}}}{\lambda _{\bf
{k}}})}\frac{\Delta _{\bf{k}}}{|\Delta_{\bf{k}}|}.
\end{equation}

The ground state is given by $|G\rangle=\prod_{\bf{k}}(u_{\bf{k}%
}+v_{\bf{k}}c_{\bf{k}\uparrow }^{\dagger}c_{-\bf{k}\downarrow}^{\dagger }
)|0\rangle $ with an energy $E_{G}$.

The impurity Hamiltonian is:

\begin{equation}
H_{I}=\epsilon _{0}\sum_{\sigma }d_{0\sigma }^{\dagger }d_{0\sigma}+U_{
0}n_{0\uparrow}n_{0\downarrow }+\sum_{\bf{k\sigma}}V_{\bf{k}%
}d_{0\sigma }^{\dagger }c_{\bf{k}\sigma }+H.c.  \label{hi}
\end{equation}
 $d_{0\sigma}^{\dagger }$ creates an electron with spin $\sigma $ in the 
impurity-
 orbital and $%
n_{0\uparrow }=d_{0\uparrow }^{\dagger }d_{0\uparrow}$.
We take the limit $U_{0}\longrightarrow \infty .$

With the Bogoliubov transformation, Eq. (\ref{bov}), the hybridization 
term (third term
in Eq. (4) is:

\begin{equation}
H_{d,c}=\sum_{\bf{k\sigma }}V_{\bf{k}}[d_{0\uparrow }^{\dagger}(u_{%
\bf{k}}^{*}\gamma _{\bf{k}\uparrow }+v_{\bf{k}}\gamma_{-\bf{%
k}\downarrow }^{\dagger })+d_{0\downarrow }^{\dagger}(-v_{\bf{k}}\gamma
_{\bf{k}\uparrow }^{\dagger }+u_{\bf{k}}^{*}\gamma_{-\bf{k}%
\downarrow })]+H.c  
\label{hdc}
\end{equation}

where we have assumed $V_{\bf{k}}=V_{-\bf{k}}$.

We also include the spin-orbit interaction between the spin of the impurity
($\bf{s}$) and the angular momentum of the conduction electrons ($%
\bf{L}$):

\begin{equation}
H_{s-O}=g\int dr^{2}\frac{\bf{s(0)\cdot L(r)}}{|\bf{r}|^{3}}
\label{hso}
\end{equation}

with $\bf{s}=d_{0\alpha }^{\dagger }\bf{\sigma }_{\alpha\beta
}d_{0\beta }$ ($\bf{\sigma }$ are the usual Pauli matrices) and ${\bf{%
L(r)}}=c_{{\bf {r}}\sigma }^{\dagger }{\bf{L}}c_{{\bf {r}}\sigma }$.

$L^{+}$ and $L^{-}$ mix states which are even with odd under
reflection about the x-y plane. As odd states have zero amplitude in the 
x-y plane $%
L^{+}$ and $L^{-}$ are irrelevant in two-dimensions, which is the case we
 consider with Copper-Oxide metals in mind. $L_{z}$ scatters a $\bf{k}$
 state to a $\bf{k}^{\bf{\prime }}$ states with $|\bf{k}|=|\bf{k}^{\prime
 }|$.

The spin-orbit interaction can therefore be rewritten as:
\begin{eqnarray*}
H_{s-O} &=&gs_{z}\sum_{\bf{k,k}^{\prime }\sigma }L_{z}^{\bf{kk}%
^{\prime }}c_{\bf{k}\sigma }^{\dagger }c_{\bf{k}^{\prime }\sigma
}=gs_{z}\sum_{\bf{k,k}^{\prime }\sigma }L_{z}^{\bf{kk}^{\prime
}}c_{%
\bf{k}\sigma }^{\dagger }c_{\bf{k}^{\prime }\sigma } \\
&=&gs_{z}\sum_{\bf{k,k}^{\prime }\sigma }L_{z}^{\bf{kk}^{\prime
}}[(\gamma _{\bf{k}\uparrow }^{\dagger }\gamma _{\bf{k}^{\prime
}\uparrow }-\gamma _{-\bf{k}\downarrow }^{\dagger }\gamma
_{-\bf{k}%
^{\prime }\downarrow })(u_{\bf{k}}u_{\bf{k}^{\prime }}^{*}+v_{%
\bf{k}}v_{\bf{k}^{\prime }}^{*})+\gamma _{\bf{k}\uparrow
}^{\dagger }\gamma _{-\bf{k}^{\prime }\downarrow }^{\dagger
}(u_{\bf{%
k}}v_{\bf{k}^{\prime }}-v_{\bf{k}}u_{\bf{k}^{\prime }})+H.c
\end{eqnarray*}

 $L_{z}$ anticommutes with the time-reversal operator (T). Therefore $L_{
z}^{%
\bf{kk}^{\prime }}=-L_{z}^{\bf{k}^{\prime }\bf{k}}=(L_{z}^{%
\bf{k}^{\prime
}\bf{k}})^{*}=iIm(L_{z}^{\bf{k}^{\prime }%
\bf{k}})$.
For planes waves $L_{z}^{(k,\varphi )(k,\varphi^{\prime
})}\simeq i%
\frac{\hbar k_{F}^{2}}{R}\sin (\varphi -\varphi^{\prime })$, where $R$ is
the radius of the sample. We absorb the coefficient in $L_{z}$ defining a
 coupling $%
 g^{\prime }.$
$L_{z}$ is invariant under rotations around $\hat{z}$ and changes sign
under reflection over planes which contains the $\hat{z}$. If the problem
has square symmetry in the x-y plane then we have a $C_{4V}$ group symmetry
and $L_{z}$ transforms as the one dimensional representation $A_{2g}$.
The other possible representation of the same group are the one dimensional
$%
A_{1g}$($s$), $B_{1g}$ ($d_{x^{2}-y^{2}}$) and $B_{2g}$ ($d_{xy}$), and
the double representation $E$ ($p_{x}$, $p_{y}$).

The model Hamiltonian considered in this paper is
\begin{equation} 
H=H_{BCS}+H_{I}+H_{s-O}
\end{equation}

\section*{Variational Calculations}

We consider a variational wavefunction  
which is a spin-doublet, reflecting
a moment in the ground state, as well as a variational wavefunction which
does not have a
moment in the ground state.

For the doublet we choose the wave-function:
\begin{equation}
|D\uparrow \rangle =(d_{0\uparrow }^{\dagger }+\sum_{\bf{k}}\alpha_{%
\bf{k}}\gamma _{\bf{k}\uparrow }^{\dagger }+d_{0\uparrow
}^{\dagger
}\sum_{\bf{kk}^{\prime }}a_{\bf{kk}^{\prime }}\gamma
_{\bf{k}%
\uparrow }^{\dagger }\gamma_{-\bf{k}^{\prime }\downarrow }^{\dagger
})|G\rangle  \label{fd}
\end{equation}
The first two terms are mixed by the hybridization, while the last term,
which is a singlet pair of quasiparticles coupled to the local moment is
mixed with the first term by the spin-orbit coupling under certain condi
tions. The $U=\infty $ constraint is obeyed.

The variational functions $\alpha _{\bf{k}}$, $a_{\bf{kk}^{\prime }}$
are determined by the condition $\delta (\langle D|H|D\rangle-E_{D}\langle
S|S\rangle )=0$, where the doublet energy ($E_{D}$) is referred to the
energy of the BCS state. We get

\begin{equation}
E_{D}=\epsilon _{0}-\sum_{\bf{k}}\alpha
_{\bf{k}}V_{\bf{k}}u_{%
\bf{k}}^{*}-\frac{g}{2}\sum_{\bf{kk}^{\prime
}}L_{z}^{\bf{kk}%
^{\prime }*}(u_{\bf{k}}^{*}v_{\bf{k}^{\prime
}}^{*}-v_{\bf{k}%
}^{*}u_{\bf{k}^{\prime }}^{*})a_{\bf{kk}^{\prime }}  \label{ed}
\end{equation}

 One should notice that $\alpha _{\bf{k}}$ and $a_{\bf{kk}%
^{\prime }}$ have
the symmetry of $V_{\bf{k}}$ and $L_{z}\cdot v_{\bf{k}}$
respectively. In a pure $d_{x^{2}-y^{2}}$ superconductor
$a_{\bf{kk}^{\prime }}$ has $d_{xy}$ symmetry 
(as $\Gamma_{A2g}\otimes \Gamma_{B1g}=\Gamma_{B2g}$).
 In the symmetric $s$-wave $a_{\bf{kk}^{\prime
}}=0 $ because the form factor
$F_{f}=u_{\bf{k}}v_{\bf{k}^{\prime
}}-v_{\bf{k}}u_{\bf{k}^{\prime }}$ vanish for $|%
\bf{k}|=|\bf{k}^{\prime }|$. It is also worth noting that momenta $k$, 
$k^{\prime }$with different signs of the coherence factor $v_{\bf{k}}$
 give the largest contributions.
The complete expresion  of $\alpha _{\bf{k}}$ and $a_{\bf{kk}%
^{\prime }}$ and $E_D$ are given in the appendix.

Consider next the state without a moment in the ground state.
$H_{BCS}$ and $H_{I}$ commutes with total spin ($S_{T}$,
impurity+conduction
electron spins) while $H_{s-O}$ does not. Then the eigenstates of $H$
cannot be classified by the total spin $S_{T}$. As a consequence the usual 
singlet solution ($S_{T}=0$%
) is replaced by a state with total spin projection z, $Sz=0$, we will
continue calling this state the singlet for simplicity.

The simplest variational wave-function with $S_z=0$ has the form 

\begin{eqnarray}
|Sz &=&0\rangle =(1+\sum_{\bf{k}}\beta _{\bf{k}}S_{\bf{k}%
}^{\dagger }+\sum_{\bf{kk}^{\prime }}M_{\bf{kk}^{\prime }}\gamma
_{%
\bf{k}\uparrow }^{\dagger }\gamma _{-\bf{k}^{\prime }\downarrow
}^{\dagger }+\sum_{\bf{kk}^{\prime }\bf{k}^{\prime \prime
}}\frac{b_{%
\bf{kk}^{\prime }\bf{k}^{\prime \prime
}}}{2}T_{\bf{k}}^{\dagger
}\gamma _{\bf{k}^{\prime }\uparrow }^{\dagger }\gamma _{-\bf{k}%
^{\prime \prime }\downarrow }^{\dagger })|G\rangle   \label{s} \\
S_{\bf{k}}^{\dagger } &=&d_{0\uparrow }^{\dagger }\gamma
_{-\bf{k}%
\downarrow }^{\dagger }-d_{0\downarrow }^{\dagger }\gamma _{\bf{k}%
\uparrow }^{\dagger }  \nonumber \\
T_{\bf{k}}^{\dagger } &=&d_{0\uparrow }^{\dagger }\gamma
_{-\bf{k}%
\downarrow }^{\dagger }+d_{0\downarrow }^{\dagger }\gamma _{\bf{k}%
\uparrow }^{\dagger }  \nonumber
\end{eqnarray}
 Again the last term can be non-zero only due to spin-orbit scattering. 
The first three terms naturally
 arise in a model with hybridization and $U=\infty $ \cite{sim}.

The complete expresion of the coefficients of Eq. (10) as well as the
 expression for the ground state energy $E_S$ are given in the
appendix. 

In a normal metal, due to the Kondo-effect, the singlet always has lower energy. 
In the limit $\epsilon _{0}<<-|V|$,
\begin{equation}
\epsilon _{0}\sim -2V^{2}\rho _{0}\ln (\frac{W}{E _{D}-E_{S}})
\end{equation}
where $\rho _{0}$ in the density of states at the Fermi level and $W$ is
the half-width band. We obtain the usual Kondo temperature 
\begin{equation}
T_{K}=E_{D}-E_{S}=W\exp(\frac{-1}{2\rho _{0}J})
\label{tk}
\end{equation}
 with $J=V^{2}/|\epsilon _{0}|$. 

For an s-wave
superconductor $(\Delta _{\bf{k}}=\Delta_0)$ in the same limit, $%
E_{S}\rightarrow \epsilon _{0}+\Delta_0$ while
$E_{D}\rightarrow
\epsilon _{0}$. As a consequence the low energy state evolves from a singlet
to the doublet when -$\epsilon _{0}/|V|$ grows \cite{sim,sato}. In both cases 
the contribution of spin-orbit term is 0. 

The low energy limit of the density of
states of a superconductor with nodes (points in 2-D, lines in 3-D) at the
Fermi level is $\rho (\omega )=\rho _{0}\omega /\Delta _{0}$. The logarithmic 
singularity leading to the Kondo-effect is replaced by a term proportional to 
$\omega ln(\omega )$ in the superconductor. An approximate analytic expresion 
of  binding energy obtained from the equations (\ref{es}, \ref{beta}) in the 
appendix is 
\begin{equation}
 \epsilon _{0} \sim - 2V^{2}\rho _{0}[\frac{E_{D}^d-E_{S}^d}{\Delta _{0}}\ln%
(\frac{| E_{D}^d-E_{S}^d |}{E_{D}^d-E_{S}^d+\omega_c})+\frac{\omega_c}%
{\Delta _{0}}+\ln (\frac{W}{\omega_c+(E _{D}^d-E_{S}^d)})]
\label{esaprox}
\end{equation}

 where we have assume a superconducting density of states linear in energy up to 
$\omega_c\sim\Delta_0$, and  constant from there to $W$, and we have aproximated
 the self-energy terms in Eq. (\ref{beta}), $-\Gamma _{\bf{1}}(\lambda _{\bf{k}%
}-E_{S})-\Gamma _{\bf{2}}(\lambda _{\bf{k}}-E_{S})$, by $E_D^d-\epsilon_0$.
 Using Eq. (\ref{tk}), it it possible to re-express Eq. (\ref{esaprox}) as
\begin{equation}
 \frac{E_{D}^d-E_{S}^d}{\Delta _{0}}\ln
(\frac{| E_{D}^d-E_{S}^d |}{\Delta _{0}}) \sim \ln (\frac{\Delta _{0}}{T_K}).
\label{dc}
\end{equation}

We conclude from equation (\ref{dc}) that the doublet has lower energy if 
$\Delta_0 > T_K$.

In Figure (1) we show $E_{D}$ and $E_{S}$ as a function of $\Delta _{0}$ for a
$d_{x^{2}-y^{2}}$ order parameter $\Delta _{\bf{k}}=\Delta _0\cos
(2\varphi )$, where $\varphi $ is the polar angle, and an  $s$-wave 
superconductor, $\Delta_{\bf{k}}=\Delta _{0}.$ We assume a constant 
non-interacting
density of states $\rho _{0}=1/2W$, with $W$ the half-band-width. Other
parameters are $\epsilon _{0}/W=-0.4$, $V/W=0.28.$ It can be seen that
while
$E_{D}$ remain almost constant and equivalent for $s$ and $d$, $E_{S}$
 grows as
$\Delta $ increases. This reflects the displacement of the
low-energy excited states to higher energies and is faster for the gapped
 $s$%
-wave ($E_{S}^{s}$). As a consequence there is a crossing of levels at a
given $\Delta _{c}.$ In the inset the effect of the spin-orbit
coupling is shown. Both $E_{D}^{d}$ and $E_{S}^{d}$ takes advantage of this term
gaining a similar amount of energy and leaving $\Delta _{c}^{d}$
unaffected.

Figure (2) shows $\Delta _{c}^{d}$ as a function of $|\epsilon _{0}|/W$ for
different values of $J$. It can be see
that $\Delta _{c}^{d}$ aproximately scales with $T_{K}=W\exp (-W/J)$. The 
quotient $\Delta _{c}^{d}/T_{K}$ lies in the range
 $0.8-2.4$ while $T_K$ and $\Delta_0$ vary by several orders of magnitude.

Changes in the macroscopic superconducting
density of states $\rho (\omega \sim 0)$ will change the above results. A 
finite concentration of impurities in a $d$ superconductor
induces a finite $\rho (\omega =0)$ \cite{haas} and helps
to stabilized the singlet ,while a complex macroscopic phase such as 
$d_{x^{2}-y^{2}}+id_{xy}$ has the opposite effect.

We can see how the doublet modifies the superconducting condensate:
\begin{equation}
\frac{\langle D\sigma |c_{-\bf{k}\downarrow }c_{\bf{k}\uparrow
}|D\sigma \rangle }{\langle D\sigma |D\sigma \rangle
}= u_{\bf{k}}v_{%
\bf{k}}(1-\frac{\alpha _{\bf{k}}^{2}+\sum_{\bf{k}^{\prime
}}a_{%
\bf{kk}^{\prime }}^{2}}{1+\sum_{\bf{k}}\alpha _{\bf{k}%
}^{2}+\sum_{\bf{kk}^{\prime }}a_{\bf{kk^{\prime }}}^{2}})
\label{fkk}
\end{equation}
It can be seen that the presence of the impurity diminishes the
($\bf{%
k\uparrow }$, $-\bf{k\downarrow }$) pairing; this reduction is of the
order 1/N, (N is the number of sites), for one impurity but is measurable for a
finite concentration of impurities. For $V_{k}=V,$ the largest values
of $%
\alpha _{\bf{k}}^{2}$ occur at the nodes of $\Delta_{\bf{k}}$
. Therefore  these states loose a greater relative weight.
This tendency agrees with the extended gapless region located in the gap
function centered around the nodes, found in a momentum-dependent 
scattering for non-magnetic impurities in
unconventional superconductors \cite{haas}. The singlet has a similar effect
on the correlation function $\langle c_{-\bf{k}\downarrow }c_{\bf{k}%
\uparrow }\rangle $ ( $\alpha _{\bf{k}}^{2}+\sum_{\bf{k}^{\prime
}}a_{\bf{kk}^{\prime }}^{2}$ should be replace by $2\beta_{\bf{k}%
}^{2}+\sum_{\bf{k}^{\prime }}M_{\bf{kk}^{\prime}}^{2}+\sum_{\bf{%
k}^{\prime }\bf{k}^{\prime \prime }}b_{\bf{kk}^{\prime}\bf{k}%
^{\prime \prime }}^{2}$ in Eq.(13) ). In this case a bigger distortion is
expected (in general $\beta _{k}$ is greater than $\alpha _{k}$).

The spin-Orbit interaction introduces a new correlation:

\begin{equation}
\langle D\sigma |D\sigma \rangle F_{\bf{kk}^{\prime }}^{D}=\langle
D\sigma |c_{-\bf{k}\downarrow }c_{\bf{k}^{\prime }\uparrow
}|D\sigma
\rangle =\frac{ig\sigma Im(L_{z}^{\bf{kk}^{\prime }})(u_{\bf{k%
}^{\prime }}v_{\bf{k}^{\prime
}}-v_{\bf{k}}u_{\bf{k}})}{(\lambda_{\bf{k}}+\lambda _{\bf{k}^{\prime }}+%
\delta _{D})}
\label{fkkp}
\end{equation}
If the symmetry of the pure superconductor is $d_{x^{2}-y^{2}}$, 
$F_{\bf{%
kk}^{\prime }}^{D}\sim id_{xy}$.  We can calculate the
induced order parameter in real space:

\begin{equation}
\Delta _{i}({\bf{R}},{\bf{r}})=\frac{\langle D\sigma
|c_{{\bf{R}}-%
{\bf{r}}\downarrow }c_{{\bf{R+r}}\uparrow }|D\sigma \rangle
}{\langle
D\sigma |D\sigma \rangle }=\frac{1}{N}\sum_{\bf{kk}^{\prime }}e^{i%
{{\bf{R(k-k}}^{\prime })}}e^{i{\bf{r(k+k}}^{\prime })}F_{\bf{kk}%
^{\prime }}^{D}
\end{equation}

Note that as $\bf{k\neq }$ $\bf{k}^{\prime },$ $\Delta _{i}$
depends on $\bf{R}$. Taking the impurity site ($\bf{R}=0$) as the center
of the symmetry this state has a $d_{xy}$ symmetry. 
Centered at a different site $\bf{R}$, $\Delta_{i}(\bf{R},\bf{r}%
)$ does not have a definite symmetry in the relative coordinate
$\bf{r}$.

This state produces a complex pattern of spontaneous
currents around the impurity which can be seen evaluating the current 
density operator:

\begin{equation}
{\bf {j}}({\bf{R}}) \langle D\sigma |D\sigma \rangle 
=\langle D\sigma |{\bf \hat{j}}({\bf{R}})|D\sigma \rangle
=\sigma c \sum_{\bf{kk}^{\prime }}e^{i%
{{\bf{R(k-k}}^{\prime })}} {\bf{(k+k}}^{\prime }) (u_{\bf{k}}v_{\bf{k}^{\prime}}%
-v_{\bf{k}}u_{\bf{k}^{\prime }})a^*_{\bf{kk}^{\prime }}
\label{j}
\end{equation}

where the constant $c=e\hbar/m Vol$. 

Expanding the form factor 
$(u_{\bf{k}}v_{\bf{k}^{\prime}}-v_{\bf{k}}u_{\bf{k}^{\prime }})^2$ of Eq. 
(\ref{j}) in the first two spherical harmonic we can get an approximate 
analytical form for this current,
\begin{equation}
{\bf {j}}({R}, \varphi) \propto \frac{J_1^2(R)}{R} -Cos^2(\varphi)%
(\frac{J_1^2(R)}{R}-8\frac{J_1(R)J_2(R)}{R^2})+...
\end{equation}
where $J_n(R)$ is the usual Bessel function of order n.

In Fig. (3) we show the current density {\bf {j}}({\bf{R}}). 
The magnitude and direction of the arrows represent the current at that point.
 This pattern can be understood as the sum of a net current around the impurity
 and eight secondary currents. Four of them are around the points 
${\bf R}k_F=(\pm 4,0)$ and ${\bf R}k_F=(0,\pm 4)$, and have the same sense as 
the net current, and the other are around the
 points ${\bf R}k_F=(\pm 4,\pm 4)$, with the oppossite sense (see Fig. (4)).
 The sense of the net current depends on the
impurity spin projection $s_{z}.$ Fig. (5) shows the net current flowing 
($J(R)=\frac{1}{2\pi}\oint {\bf j}({\bf R }){ \bf d\varphi}$) as a function of
 the distance to the impurity. As before we can get the approximate analytical 
results,
\begin{equation}
J(R) \propto \frac{J_1^2(R)}{2R} +4\frac{J_1(R)J_2(R)}{R^2}+...
\end{equation} 

 The sign of the current varies with distance but the total current around the 
impurity is not zero. We have neglected any diamagnetic current produced by the 
magnetic field induced by the spontaneous current. 

For the singlet the two momenta
correlation gives $F_{\bf{kk}^{\prime }}^{S}=M_{\bf{kk}^{\prime
}}(u_{\bf{k}^{\prime }}u_{\bf{k}}-v_{\bf{k}}v_{\bf{k}%
^{\prime }})$, but for $v_{\bf{k}}$ real (real macroscopic
superconducting state), this state preserves the time-reversal symmetry
($%
T|S\rangle =|S\rangle $) and does not have any spontaneous current.

\section*{Conclusions}

In summary we have shown, in agreement with previous results, that a magnetic 
impurity in a $d_{x^{2}-y^{2}}$ superconductor has a transition from a Kondo 
singlet to an unscreened doublet as the superconductivity is turned on. 
The coupling between the spin of the impurity and the orbital momenta of the 
electrons induces a complex secondary component of the superconducting order
 parameter around the impurity in the doublet.

What are the observable consequences of the complex order parameter around the
 magnetic impurity for the case of dilute impurity concentrations? 
A complex order parameter by itself leads to a finite gap in the excitation
spectra of the supeconductor. But the potential scattering due to the impurity,
 not considered in this paper, if it is not sitting in a center of symmetry of 
the crystal, produces a finite density of states at low energies and spoils this 
effect. 
The principal effect of such a potential scattering in the current around the 
impurity is that it decays in a length of the order of magnitude of the 
mean-free path, rather than decay as a power law obteined in this paper.

Global time reversal breaking due to the current around the magnetic impurities
 is possible only if they are aligned ferromagnetically. 
This is because the direction of the current depends on the direction of the 
magnetic moment.
 Ferromagnetic aligment is unlikely in general.
 A spin glass state, with an associated glassy pattern of currents, 
is much more likely. 
If a ferromagnetic aligment of the impurity spins is achieved, the nature of 
superconductivity is strongly affected and may be even destroyed. 
The physical properties in such a case have been discussed for s-wave 
superconductors \cite{blo,ng}

\section*{acknowledgments}
M.E.S.  would like to thank L. Balents for useful discussions. 
M.E.S is a fellow of the Consejo Nacional de Investigaciones 
Cient\'{\i}ficas y t\'ecnicas (CONICET), Argentina.

\section*{appendix A}

The coefficient of the doublet (Eq. \ref{fd}) are given by:
\begin{eqnarray}
a_{\bf{kk}^{\prime }} &=&-i\frac{g}{2l_{\bf{kk}^{\prime
}}}Im%
(L_{z}^{\bf{kk}^{\prime }})(u_{\bf{k}}v_{\bf{k}^{\prime }}-v_{%
\bf{k}}u_{\bf{k}^{\prime }})+\frac{V_{\bf{k}}v_{\bf{k}}V_{%
\bf{k}^{\prime }}v_{\bf{k}^{\prime }}^{*}}{(\lambda_{\bf{k}%
}-E_{D})l_{\bf{kk}^{\prime
}}}+\frac{V_{\bf{k}}v_{\bf{k}}^{*}}{%
(\lambda _{\bf{k}}-E_{D})l_{\bf{kk}^{\prime }}}\sum_{\bf{k}%
^{\prime \prime }}V_{\bf{k}^{\prime \prime }}v_{\bf{k}^{\prime
\prime }}a_{\bf{k}^{\prime \prime }\bf{k}}  \label{akk} \\
l_{\bf{kk}^{\prime }} &=&\epsilon _{0}+\lambda _{\bf{k}}+\lambda_{%
\bf{k}^{\prime }}-E_{D}  \nonumber \\
\alpha _{\bf{k}} &=&\frac{-V_{\bf{k}}u_{\bf{k}}}{\lambda _{%
\bf{k}}-E_{D}-\Gamma _{a\bf{k}}}-i\frac{g}{2(\lambda_{\bf{k}%
}-E_{D}-\Gamma _{a\bf{k}})}\sum_{\bf{k}^{\prime }}V_{\bf{k}%
^{\prime }}v_{\bf{k}^{\prime }}^{*}Im(L_{z}^{\bf{kk}^{\prime
}})\frac{(u_{\bf{k}}v_{\bf{k}^{\prime }}-v_{\bf{k}}u_{\bf{k}%
^{\prime }})}{l_{\bf{kk}^{\prime }}}  \label{alfak} \\
\Gamma _{a\bf{k}} &=&\sum_{\bf{k}^{\prime }}\frac{V_{\bf{k}%
^{\prime }}^{2}v_{\bf{k}^{\prime }}^{2}}{\epsilon _{0}+\lambda_{\bf{%
k}}+\lambda _{\bf{k}^{\prime }}-E_{D}}  \nonumber
\end{eqnarray}

The second terms in Eqs. \ref{akk} and \ref{alfak} ($\alpha_{\bf{k}}$
and $a_{\bf{kk}^{\prime }}$) only gives a contribution to the energy if $
\Gamma _{V}\otimes \Gamma _{v}\otimes \Gamma _{Lz}=\Gamma _{V}$,
where $\Gamma _{V}$, $\Gamma _{v}$ and $\Gamma _{Lz}$ are the respective
irreducible representations. In a $C_{4V}$ group this is only possible if
 $%
\Gamma _{V}=E$ ($p_{x}$, $p_{y}$). This means that the excited (second 
and
third terms in Eq. \ref{akk}) connected with $d_{0\uparrow }^{\dagger }$ 
via
$H$, are not connected to each other.

Considering the most relevant terms the doublet energy $E_D$ is given by
\begin{equation}
 E_{D}-\epsilon
_{0}=-\delta_{D}=-\Gamma_{\bf{1}}(-E_{D})-\Gamma_{\bf{2}%
}(-E_{D})
\end{equation}

 with:

\begin{eqnarray*}
\Gamma_{\bf{1}}(x) &=&\sum_{\bf{k}^{\prime}}\frac{V_{\bf{k}%
^{\prime }}^{2}u_{\bf{k}^{\prime }}^{2}}{\epsilon_{0}+\lambda_{\bf{%
k}^{\prime }}+x} \\
\Gamma_{\bf{2}}(x) &=&\frac{g^{2}}{4}\sum_{\bf{kk}^{\prime
}}\frac{%
|L_{z}(uv-vu)|_{\bf{kk}^{\prime }}^{2}}{\epsilon_{0}+\lambda_{\bf{k%
}^{\prime}}+\lambda_{\bf{k}}+x}
\end{eqnarray*}

\section*{appendix B}

For the singlet (Eq. \ref{s}), the ground state energy is calculated to 
be:
\begin{equation}
E_{S}=\epsilon _{0}-\delta _{S}=2\sum_{\bf{k}}\beta
_{\bf{k}}V_{%
\bf{k}}v_{\bf{k}}^{*} \label{es}
\end{equation}
where
\begin{equation}
\beta _{\bf{k}}=\frac{-V_{\bf{k}}v_{\bf{k}}}{\lambda_{\bf{k}%
}+\epsilon _{0}-E_{S}-\Gamma _{\bf{1}}(\lambda _{\bf{k}%
}-E_{S})-\Gamma _{\bf{2}}(\lambda _{\bf{k}}-E_{S})}+\beta_{\bf{k%
}}^{corr}
\label{beta}
\end{equation}
The coefficients are given by

\begin{eqnarray}
\beta _{\bf{k}}^{cor} &=&\frac{\sum_{\bf{k}^{\prime }}\beta _{%
\bf{k}^{\prime }}(A_{\bf{kk}^{\prime }}^{(1)}-A_{\bf{kk}^{\prime
}}^{(2)}+ig/2(L_{z}(uu^{*}+vv^{*})_{\bf{kk}^{\prime }})}{\lambda _{%
\bf{k}}+\epsilon _{0}-E_{S}-\Gamma _{\bf{1}}(\lambda_{\bf{k}%
}-E_{S})-\Gamma _{\bf{2}}(\lambda _{\bf{k}}-E_{S})} \label{bkcorr} \\
A_{\bf{kk}^{\prime }}^{(1)} &=&\frac{V_{\bf{k}}u_{\bf{k}}V_{%
\bf{k}^{\prime }}u_{\bf{k}^{\prime }}}{\lambda_{\bf{k}}+\lambda
_{\bf{k}^{\prime }}-E_{S}}  \nonumber \\
A_{\bf{kk}^{\prime }}^{(2)} &=&\frac{g^{2}}{4}\sum_{\bf{k}^{\prime
\prime }}\frac{L_{z}(uv-vu)_{\bf{k}^{\prime }\bf{k}^{\prime \prime
}}L_{z}(uv-vu)_{\bf{kk}^{\prime \prime }}^{*}}{\lambda _{\bf{k}%
}+\lambda _{\bf{k}^{\prime }}+\lambda _{\bf{k}^{\prime \prime
}}+\epsilon _{0}-E_{S}}  \nonumber \\
M_{\bf{kk}^{\prime }} &=&\frac{-(V_{\bf{k}}u_{\bf{k}}\beta_{%
\bf{k}^{\prime }}+V_{\bf{k}^{\prime }}u_{\bf{k}^{\prime
}}\beta_{\bf{k}})}{\lambda _{\bf{k}}+\lambda _{\bf{k}^{\prime
}}-E_{S}} \\
b_{\bf{kk}^{\prime }\bf{k}^{\prime \prime }}
&=&-i\frac{g}{2}\frac{%
\beta _{\bf{k}}Im(L_{z})(uv-vu)_{\bf{k}^{\prime }\bf{k}%
^{\prime \prime }}}{\lambda _{\bf{k}}+\lambda _{\bf{k}^{\prime
}}+\lambda _{\bf{k}^{\prime \prime }}+\epsilon _{0}-E_{S}}
\end{eqnarray}

$A_{\bf{kk}^{\prime }}^{(1)},$ $A_{\bf{kk}^{\prime }}^{(2)}$ and the
last term in Eq. \ref{bkcorr} contribute only in very special cases 
(i.e. $A_{\bf{kk}^{\prime }}^{(1)}$  gives a contribution only if 
$\Gamma _{V}\otimes \Gamma _{V}=\Gamma _{v}$
 ($\Gamma _{V}=\Gamma _{v}=A_{1g} (s)$ 
or $ \Gamma _{V}=E (p_x, p_y)$). 
They do not contribute for the  d-wave superconductor consider in the main text.

\section*{Captions}
  
{\bf Figure (1)} - Doublet energy, $E^s_{D}=E_D^d$ (full line), and singlet
 energies $E_{S}^s$ (dotted line) and $E_{S}^d$ (dashed line) as a function
 of $\Delta _{0}$, for  $s$-wave and $d_{x^{2}-y^{2}}$  superconductors. 
 $\Delta^s_{c}$ and  $\Delta^d_{c}$ denote the  values of the superconducting 
order parameter where the levels crosses. Other parameters are 
$\epsilon _{0}/W=-0.4$, $V/W=0.28.$.  $\Delta _{c}.$ 
The inset shows $E_{D}^{d}$ and $E_{S}^{d}$ with the inclusion of  an 
spin-orbit interaction, with a constant coupling $g'/W=.1$, (dashed lines) 
compare with the $g'=0$ case (full lines), near the crossing $\Delta^d_{c}$.

{\bf Figure (2)} Critical values $\Delta _{c}^{d}/T_K(J)$ as a function of 
$|\epsilon _{0}|/W$ for different values of $J$.

{\bf Figure (3)} Current density  ${\bf j}({\bf {R}})$ around the impurity. 
The magnitude and direction of the arrows represent the current at that point.
 The sense of the currents depends on the impurity spin projection $s_{z}.$
 
{\bf Figure (4)} Current density  ${\bf j}({\bf {R}})$ around the impurity 
after substracting the net current (Fig. 5). The magnitude and direction of 
the arrows represent the current at that point.

{\bf Figure (5)} Net current $J(R)=1/2\pi\oint {\bf j}({\bf R}) { \bf d\varphi}$
 as a function of the distance to the impurity. 

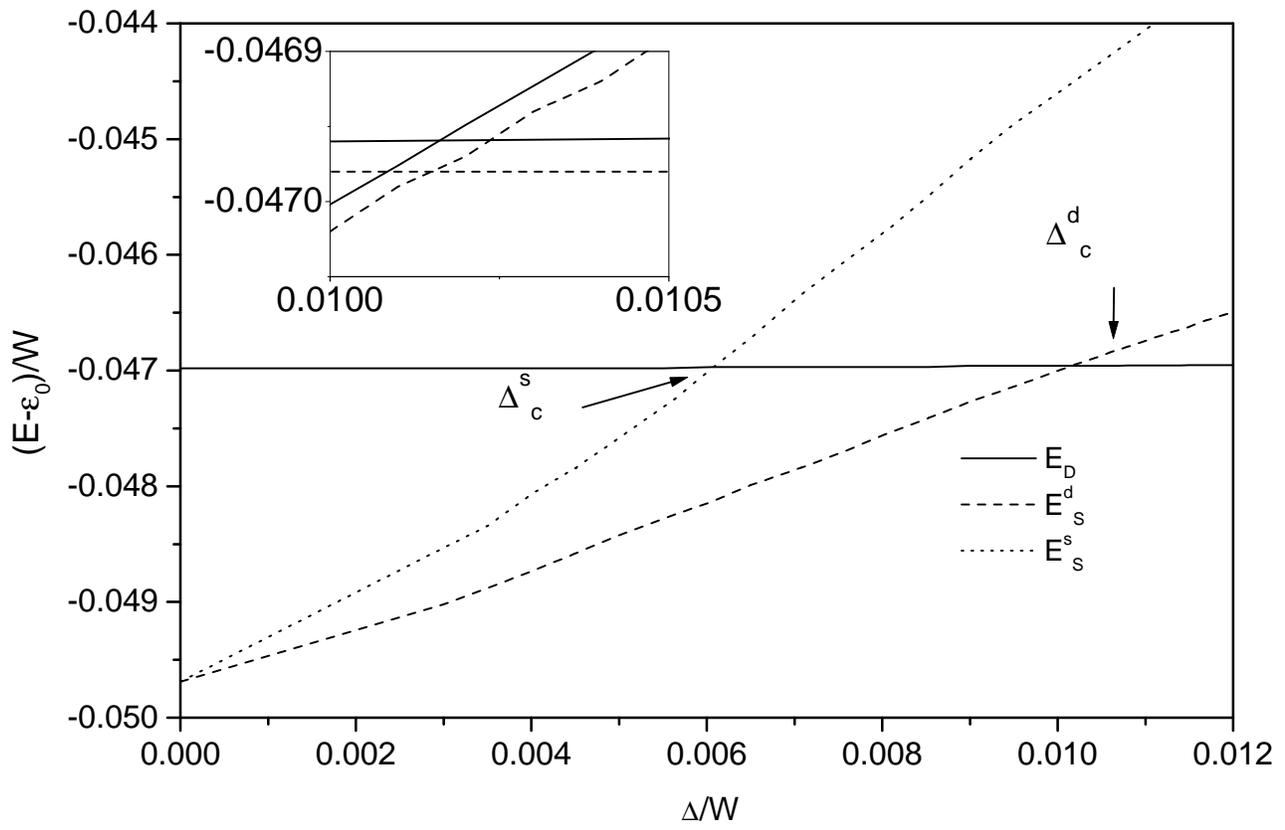
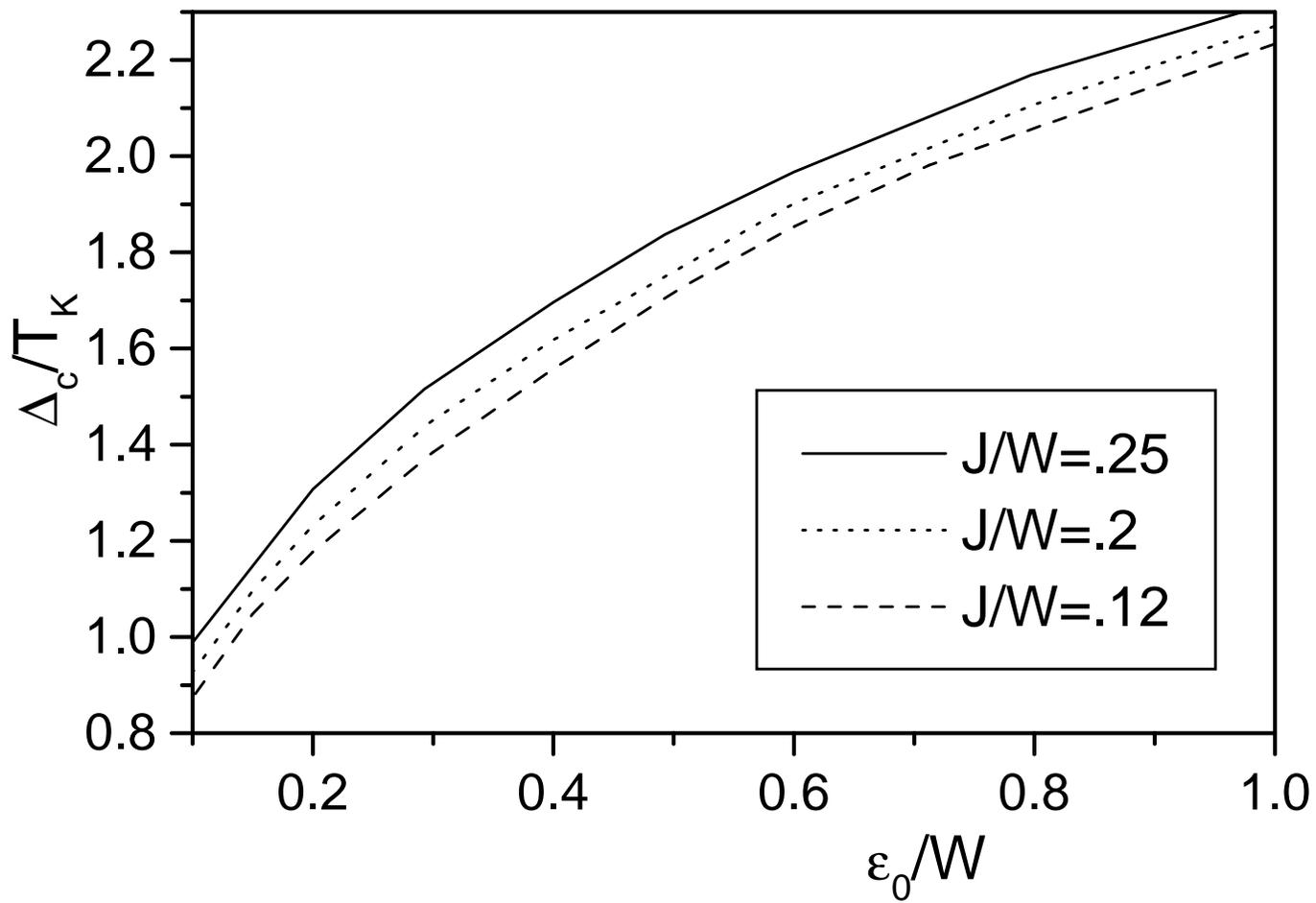
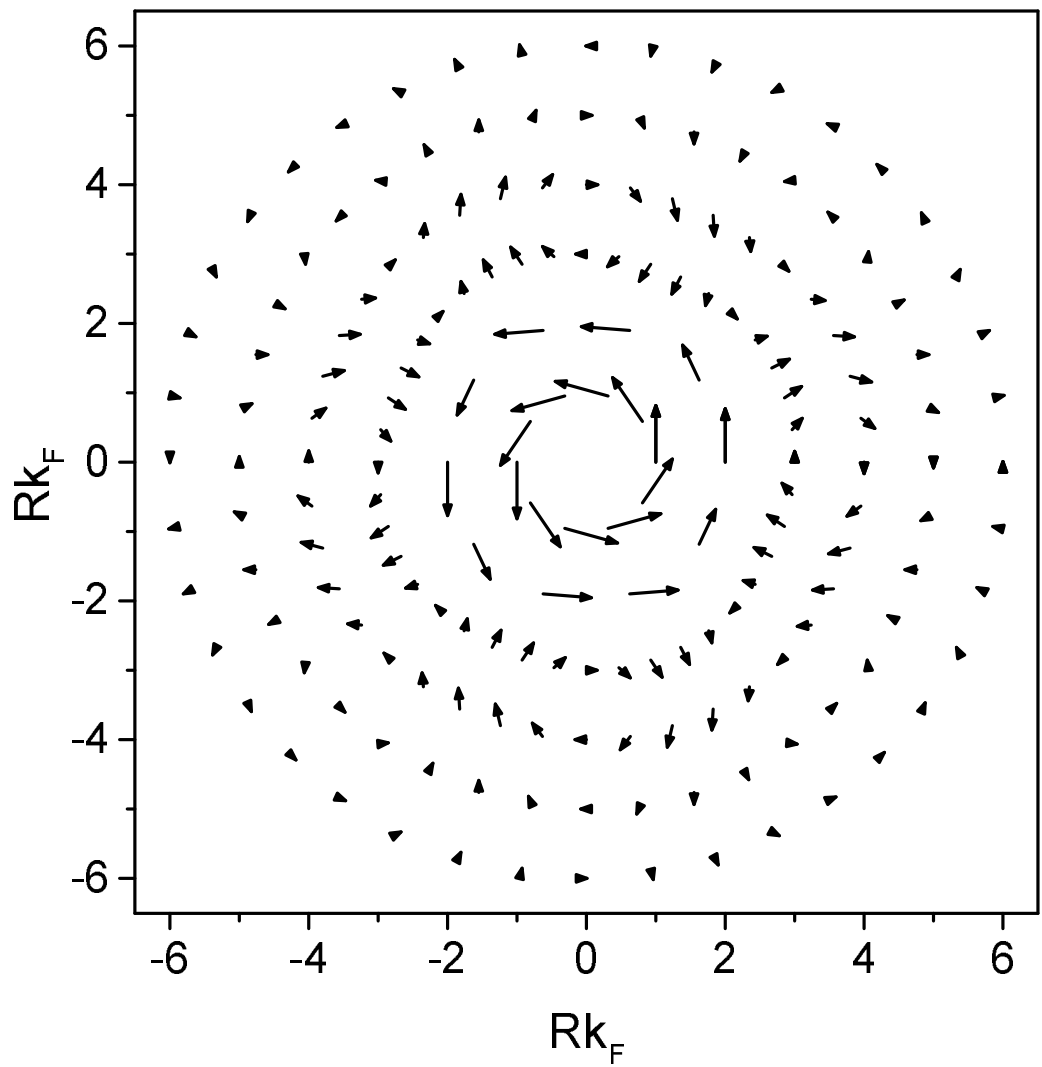
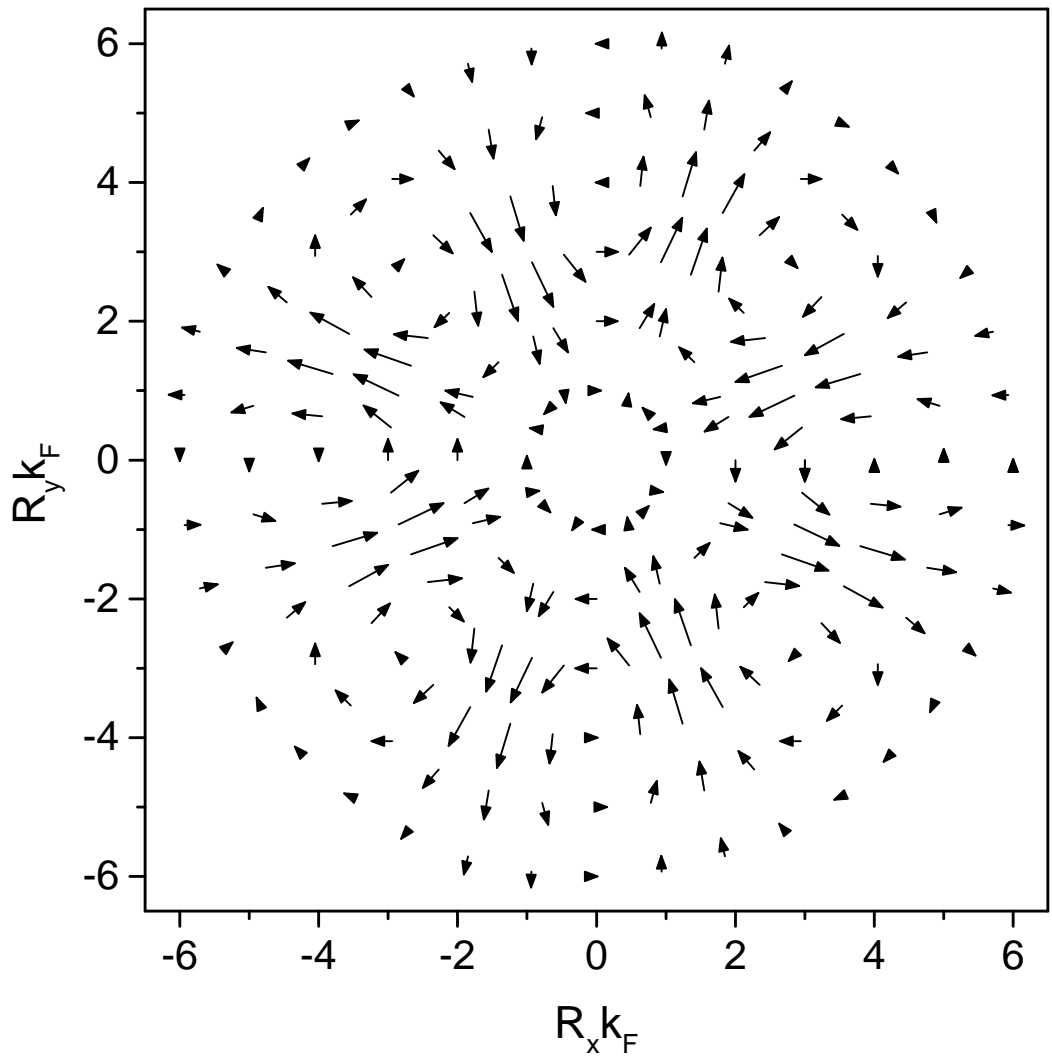
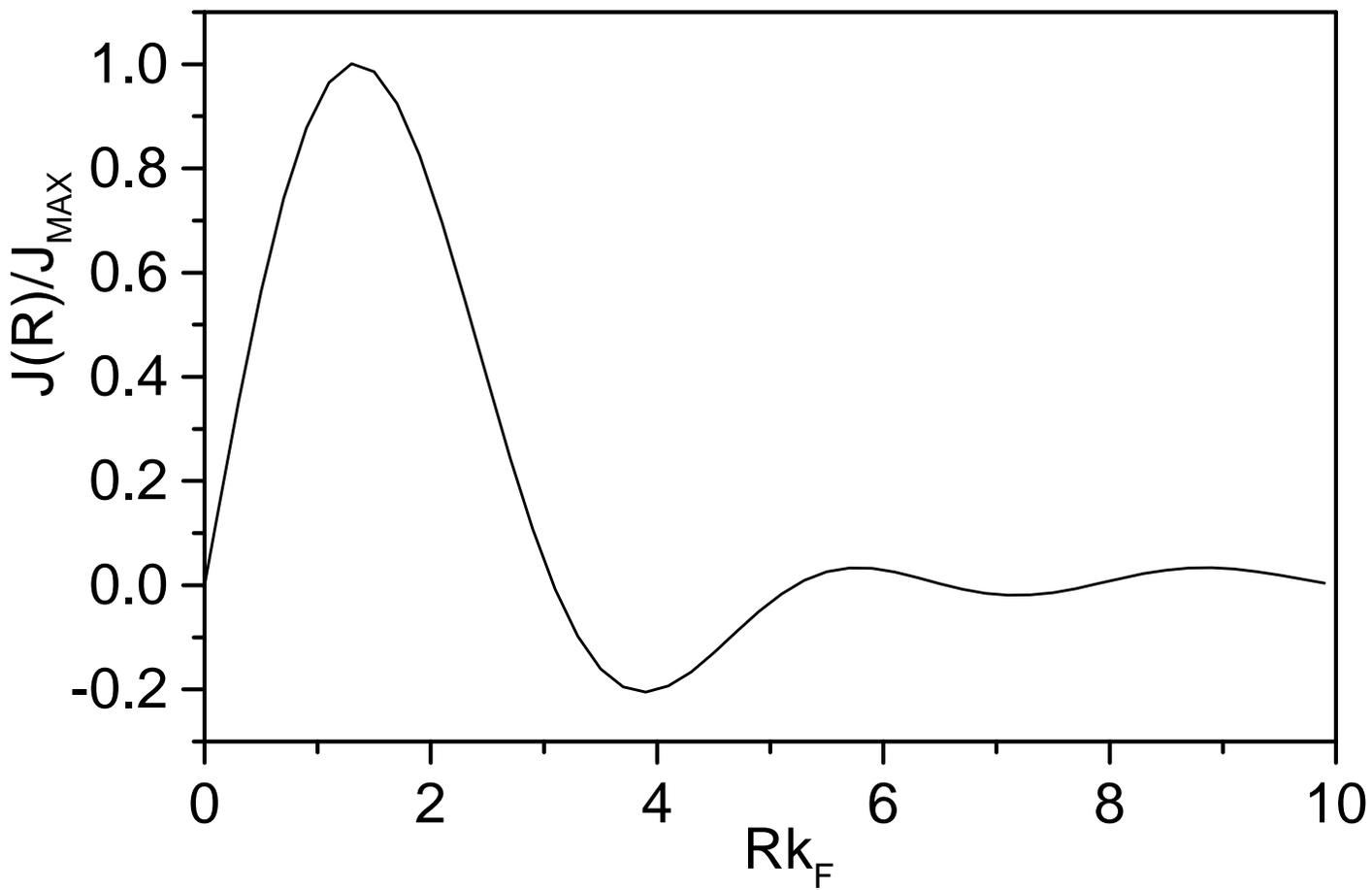
\end{document}